\documentclass[10pt,aps,prl,twocolumn,superscriptaddress,groupedaddress]{revtex4-1}

\usepackage{graphicx}
\usepackage{dcolumn}
\usepackage{bm}
\usepackage{amssymb}
\usepackage{amsmath}
\usepackage{float}
\usepackage{epsfig}
\usepackage{amsmath}
\usepackage{color}
\usepackage{txfonts}
\usepackage{microtype}
\usepackage{ulem}
\usepackage{pdfpages}

\makeatletter
\newcommand*{\rom}[1]{\expandafter\@slowromancap\romannumeral #1@}
\AtBeginDocument{\let\LS@rot\@undefined}
\makeatother

\begin{document}

\title{Nonlinear QED in an ultrastrong rotating electric field: Signatures of \\the momentum-dependent effective mass}


\author{E. Raicher\footnote{E-mail address: erez.raicher@mpi-hd.mpg.de }}
\affiliation{Max-Planck-Institut f\"{u}r Kernphysik, Saupfercheckweg 1,  69117 Heidelberg, Germany }
\author{K. Z. Hatsagortsyan}
\affiliation{Max-Planck-Institut f\"{u}r Kernphysik, Saupfercheckweg 1,  69117 Heidelberg, Germany }

\date{\today}

\begin{abstract}

The specific features of nonlinear pair production and radiation processes in an ultratsrong rotating electric field are investigated, taking into account that this field models the antinodes of counterpropagating laser beams. It is shown that a particle in a rotating electric field acquires an effective mass which depends on its momentum absolute value as well as on its direction with respect to the field plane. This phenomenon has an impact on the nonlinear Breit-Wheeler and nonlinear Compton processes. 
The spectra of the produced pairs in the first case, and the emitted photon in the second case, are shown to bear signatures of the effective mass.
In the first case, the threshold for pair production by a $\gamma$-photon in the presence of this field varies according to the photon propagation direction. In the second case, varying the energy of the incoming electron allows for the measurement of the momentum dependence of the effective mass. 
Two corresponding experimental setups are suggested.

\end{abstract}

\maketitle

A strong field may modify the mass of the particles with which it interacts.
This phenomenon, originally introduced in the context of particle physics (the Higgs mechanism \cite{Higgs}), may be also found in condensed matter \cite{landau2}, plasma \cite{Mendonca}, and strong field QED \cite{ritus1,kibble,landau}. In the latter, the effective mass significantly deviates from the vacuum mass for 
large values of the classical strong field parameter 
$\xi \equiv ea/m $ \cite{ritus}, where $a$
is the amplitude of the laser vector potential $A_{\mu}$, and
$-e$ and $m$ are the electron charge and mass, respectively; relativistic units $\hbar = c = 1$ are used.
Contemporary optical lasers \cite{yoon,gemini} may reach  $\xi \sim 100$ and a significant increase is expected in the next generation laser facilities \cite{ELI, XCELS}. Consequently, the effective mass is expected to play a significant role in the interaction of such intense beams with matter.

In the realm of the strong field QED, the perturbation treatment is developed 
based on solutions of the Dirac equation in the presence of the external field \cite{Furry}. 
The fundamental quantity of this theory is the quantum strong field parameter 
$\chi \equiv e\sqrt{-(F^{\mu \nu} P_{\nu})^2}/m^3$ \cite{ritus}, where $P_{\nu} = (\mathcal{E}, \textbf{P})$ is the kinetic four-momentum, a bold letter stands for a 3-vector and $F_{\mu \nu} = {\partial}_\mu A_{\nu} - {\partial}_\nu A_{\mu}$ is the field tensor. 
In the limit $\chi \rightarrow 0$ the classical electrodynamic is recovered.
The lowest order processes described by this theory are \cite{ritus,DiPiazza} non-linear Compton (NLC), where an electron absorbs $s$-laser photons to emit an energetic photon, 
nonlinear Breit-Wheeler (NLBW), where an electron positron pair is created following the absorption of a $\gamma$-photon and $s$-laser photons,
and the Schwinger mechanism \cite{schwinger}, where the strong field induces a pair creation from the vacuum.
Due to the dressing of the electron mass by a strong laser field,
the kinematic associated with these processes is modified with respect to the weak field case. In particular, one may show \cite{ritus} that the quantity appearing in the energy-momentum conservation is the cycle-averaged momentum $\bar{P}_{\mu}$.
The NLC and NLBW processes are a crucial part in the physical picture of the interaction of high intensity laser with matter, see e.g. \cite{kirk_2008,sokolov_prl,fedotov,nerush,nakamura,sergeev,gonoskov}.
For a long time, the only experimental investigation of these processes was the E-144 experiment carried out in SLAC \cite{E144a, E144b}, where the effective mass was not directly observed. Recently, however, several experiments aiming at strong field QED were reported \cite{sarri,poder, Cole, wistisen}, 
bringing closer the perspective to measure the effective mass.

Theoretical investigation of the effective mass requires a solution for the dynamics of the particle in the presence of the field. It is well known that for a plane wave  field (PWF) the Dirac equation admits an analytical solution \cite{Volkov}. Owing to its high relevance to laser matter experiment, most of the existing literature concerning the effective mass relies on this solution. 
It was shown to depend on the laser polarization \cite{ritus} and the shape  of the laser pulse  \cite{Harvey}, and leaves signatures in the radiation spectrum \cite{Harvey2}. The definition of the effective mass becomes more
elusive for non-periodic fields such as few cycles pulses \cite{boca, Nousch, Heinzl, MackDip, Seipt}. 
Another field configuration bearing significance to laser matter interaction is the oscillating electric field. This field models the antinode of a standing wave, created by two counterpropagating laser beams. Note that the electrons are expected to be trapped in the antinodes of the standing laser wave in the anomalous radiative trapping regime \cite{gonoskov2014}.
Several approximations to the corresponding wave function were discussed in the context of various strong field processes \cite{brezin,Noga,becker, mocken,granz, Varro,my1,my3,Heinzl_sol,my4,my5,Mack}. The effective mass and its consequences, however, were explicitly considered only for the limiting case of a vanishing particle momentum \cite{alkofer}.

In this letter, the role of the effective mass for nonlinear QED in a strong rotating electric field (REF) is investigated. We derive the analytic expression for the effective mass of a particle in the presence of REF and  
show that it depends not only on the field parameters but also on the particle momentum.  
Namely, two particles propagating in different direction or velocity in the same field acquire a different mass.
The effect of the dressed mass on the probabilities of NLC and NLBW processes, and on the spectra of photons or created electron-positron pairs are explored by analytical and numerical means. Furthermore, two experimental scenarios are suggested to detect a measurable signature of this phenomenon.

\begin{figure}
  \begin{center}     	   
  \includegraphics[width=0.4\textwidth]{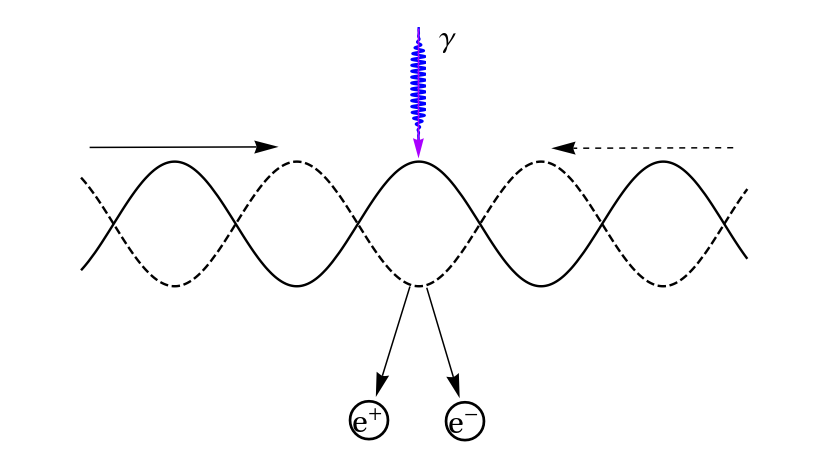}
      \caption{The schematic set up. The counterpropagating beams create a standing wave and the $\gamma$-photon beam passes through the antinode and creates electron-positron pairs.}
        \label{setP}
  \end{center}
\end{figure}

A possible realization of REF in laboratory may be achieved using counterpropagating circularly polarized laser beams, as illustrated schematically in Fig.~\ref{setP}. In the antinodes of the standing wave created by the beams the magnetic components of the two beams cancel each other and the field can be approximated as  REF. 

\begin{figure}
  \begin{center}     	   
  \includegraphics[width=0.45\textwidth]{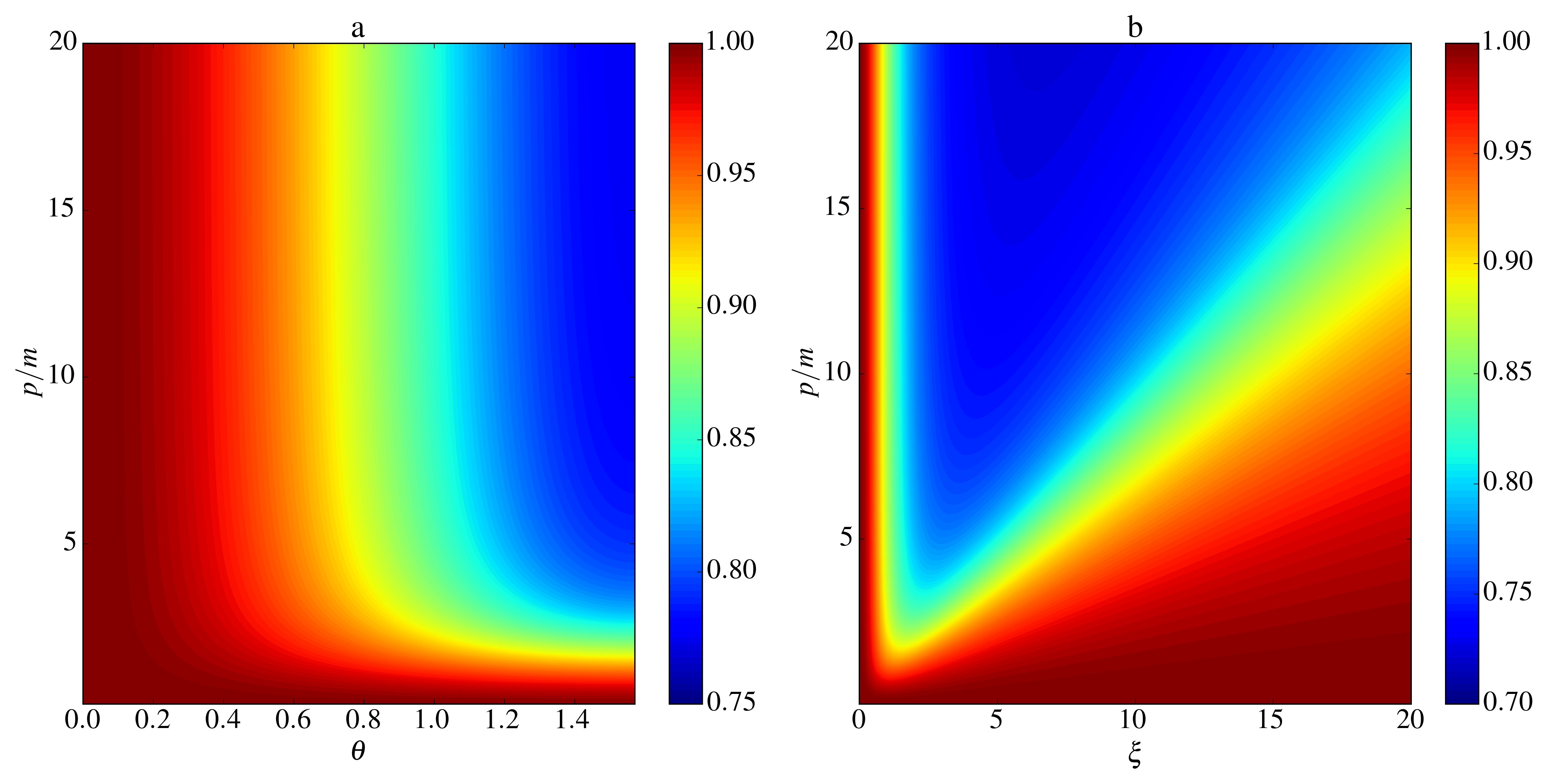}
      \caption{The effective mass $m_*$, normalized to the PWF value $m_*^P$: (a) as a function of $p/m$ and $\theta$ for $\xi=2$; (b) as a function of $p/m$ and $\xi$ for $\theta = \pi/2$.}
        \label{effMassP}
  \end{center}
\end{figure}

Firstly, let us explicitly calculate the effective mass of a particle in the field under consideration. The effective mass is defined via the cycle-averaged momentum of the electron in this field: $m_* \equiv \sqrt{\bar{P}^2}$. The vector potential of REF is defined as
$A^{\mu} = a^{\mu}_1 \cos ( \omega t) + a^{\mu}_2 \sin ( \omega t)$,
where $a^{\mu}_1 = a(0,1,0,0)$, $a^{\mu}_2 = a(0,0,1,0)$ are the polarization vectors, and the ($x$-$y$)-plane is the polarization plane. 
The time dependent momentum reads $\textbf{P}=\textbf{p}-e \textbf{A}$, with the initial momentum $p_{\mu}=(\varepsilon,\textbf{p})$, and the time dependent energy is derived from the free electron dispersion relation  $\mathcal{E} = \sqrt{m^2 + \textbf{P}^2}$.
Without loss of generality we assume that the particle propagates in the ($x$-$z$)-plane, so that $\textbf{p} = p(\sin \theta,0, \cos \theta)$, where $p \equiv |\textbf{p}|$ and $\theta$ is the angle between $\textbf{p}$ and the $z$-axis, transverse to the polarization plane. 
The cycle-average energy is given by $\bar{\mathcal{E}} = 2GE_2(\mu)/ \pi $ \cite{mocken}, where $E_2$ is the elliptic integral of the second kind, $\mu \equiv 4m \xi p |\sin \theta| /G^2$
and $G \equiv [m^2(1+\xi^2)+ p^2 + 2m \xi p |\sin \theta|]^{1/2}$. 
Generally speaking, the effective mass depends on 3 quantities: $\xi$, $p$ and $\theta$. In the following we examine analytically its limits.
For a particle initially propagating perpendicular to the field plane, i.e. $\theta = 0$, as well as for $p = 0$, one obtains $\mu = 0$. Consequently, since $E_2(0)=\pi/2$, the effective mass recovers its PWF value $m_* = m^{P}_*\equiv m \sqrt{1+\xi^2}$. 
It coincides with the result of \cite{alkofer}, which was obtained for vanishing momentum.
An explanation to this fact is suggested later on. 
In the case of $ p \ll m \xi$, one obtains $\mu \ll 1$. Since the first order Taylor expansion of $E_2$ with respect to $\mu$ vanishes, this limit corresponds, up to second order, to $m_* \approx m_*^P$.
For the opposite case ($p  \gg m \xi$), one may expand $G$ and $E_2$ \cite{Supp} appearing in the general expression, which leads to 
\begin{equation}
m_* \approx m^{P}_* \sqrt{1-\frac{\xi^2}{2(1+\xi^2)} \sin^2 \theta}. 
\label{eq:av effMass}
\end{equation}
Accordingly, the minimal value of $m_*$, corresponding to $\xi \gg 1$, is $m_* \approx m^{P}_* / \sqrt{2}$. In this limit, however, the local crossed field approximation sets in for the NLC and NLBW processes, when the probabilities and spectra depend solely on the parameter $\chi$, but not on $\xi$, and all signatures of the effective mass vanish.
Thus, the preferable range for the study of the effective mass influence is $\xi \sim 1$.
Fig. \ref{effMassP}(a) shows the effective mass for $\xi=2$ (normalized to the PWF value $m_*^P$) 
as a function of  $\theta$ and $p/m$. One may observe that for $\theta =0$ or $p \ll m \xi$ the normalized value of the effective mass tends to 1, in agreement with the analytical result.
The values for $p \gg m \xi$ coincides to a very good approximation with Eq. (\ref{eq:av effMass}). 
Fig. \ref{effMassP}(b) presents the same quantity as a function of $p/m$ and $\xi$ for $\theta= \pi/2$. 
The limits of $p$ much higher / lower than $m \xi$ hold here as in Fig. \ref{effMassP}(a). Notice that the minimal value of the normalized effective mass is $1/\sqrt{2} \approx 0.71$ and that a significant decrease appears for $p \gg m \xi, \xi \sim 1$, in accordance with Eq. (\ref{eq:av effMass}). 

Since the effective mass is embedded in the kinematics associated with the NLC and NLBW processes, its fingerprint may be found in the corresponding spectra. In a previous work \cite{my5} we have examined in details the NLC probability for this field configuration. It was found that as long as $\varepsilon \gg m \xi $, the rate coincides to an excellent approximation with the one obtained with the semiclassical formula introduced by Baier and Katkov \cite{katkov1, katkov2}. Due to the crossing symmetry between the matrix elements of the Compton and Breit-Wheeler processes \cite{peskin}, 
this conclusion holds for the NBW process as well. For this reason, we calculate here the rate according to the semiclassical expression. In this case,
the probability to emit a photon with a four-momentum $k'=(\omega',\textbf{k}')$ reads
\begin{equation}
d \mathcal{P} = \frac{\alpha}{(2 \pi)^2 \omega'} |\mathcal{M}|^2 d^3\textbf{k}',
\label{eq:av P}
\end{equation}
where $\alpha \approx 1/137$ is the fine structure constant,	 
\begin{equation}
|\mathcal{M}|^2 \equiv
-\frac{\varepsilon'^2 + \varepsilon^2 }{2  \varepsilon'^2}
|\mathcal{T}_{\mu}|^2 + 
\frac{m ^2 \omega'^2}{2 \varepsilon'^2 \varepsilon'^2}
|\mathcal{I}|^2,
\label{eq:av K}
\end{equation}
and $\varepsilon' =  \varepsilon - \omega'$. The integrals $\mathcal{I}$ and $\mathcal{T}_{\mu}$ are defined as follows
\begin{equation}
\mathcal{I} \equiv \int_{-\infty}^{\infty}{dt}  e^{i \psi}, \quad \quad \mathcal{T}_{\mu} \equiv \int_{-\infty}^{\infty}{dt} \upsilon_{\mu} e^{i \psi},
\label{eq:av TI}
\end{equation}
where the phase reads $\psi \equiv \frac{\varepsilon}{\varepsilon'} k' \cdot x(t)
$, the velocity is $\upsilon_{\mu}=P_\mu/ \mathcal{E}$ and $x(t)$ designates the classical trajectory. The probability associated with the NLBW takes analogous form where $d^3\textbf{k}'$ is replaced by the momentum of the outcoming electron $d^3\textbf{p}'$ and $\varepsilon' = \omega'-\varepsilon$.
It follows from Eqs.~(\ref{eq:av P})-(\ref{eq:av TI}) that the probability is determined according to the trajectory of the electron in the presence of the field. It provides an explanation to the fact that for $\theta = 0$ the effective mass coincides with that of the PWF, as seen from Eq.~(\ref{eq:av effMass}). 
In this case, the particle is simply moving in a circle in the ($x$-$y $)-plane while drifting along the $z$-axis, which is identical to the particle motion in a PWF. 

\begin{figure}
  \begin{center}     	   \includegraphics[width=0.45\textwidth]{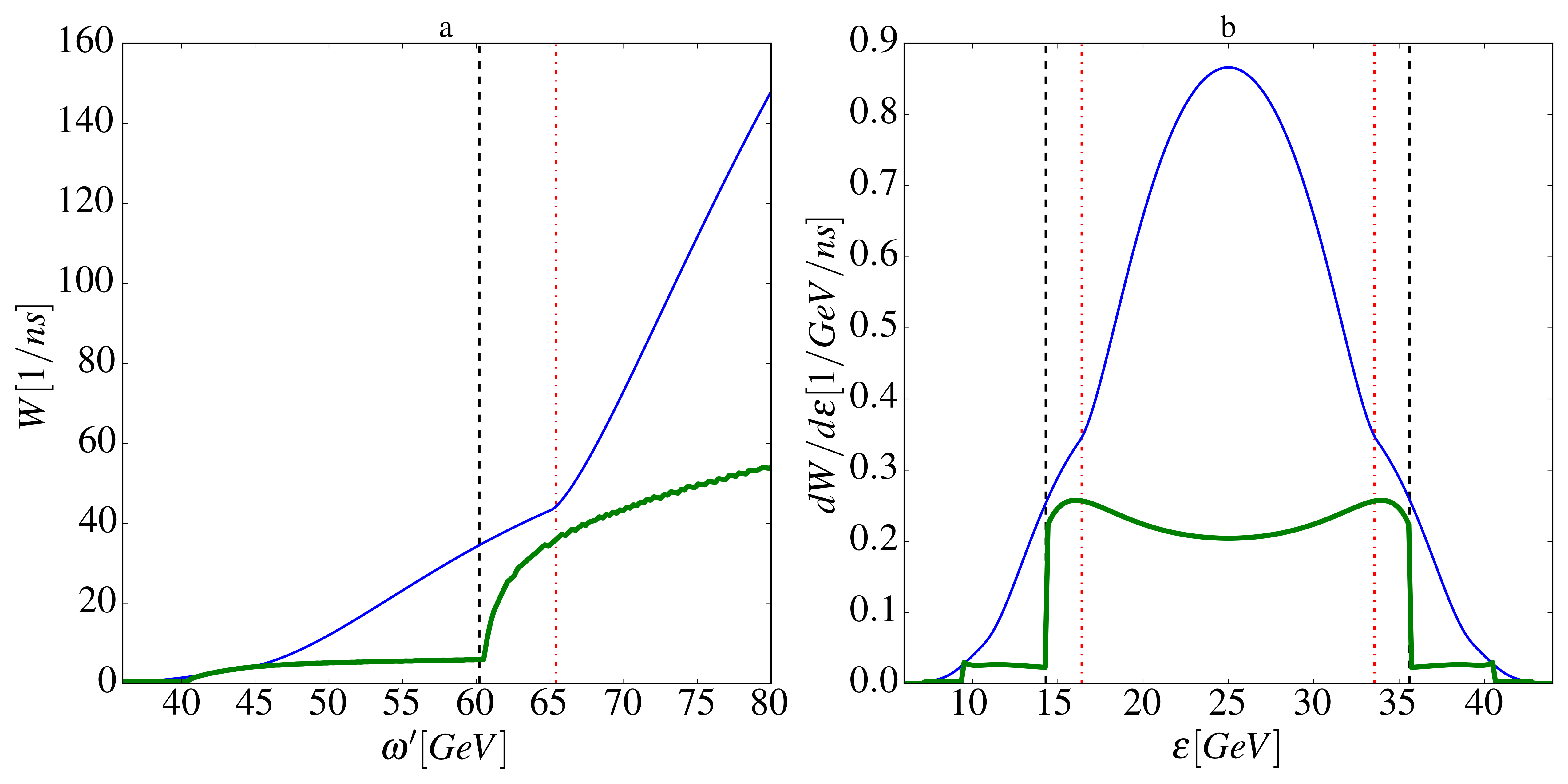}
      \caption{(a) Total pair production probability vs. incoming $\gamma$-photon energy: $\theta=0$ (thin blue line) and $\theta=\pi/2$ (green line). The vertical dash-dotted red (dashed black) line shows the threshold location $\omega'_2$ for $\theta = 0$ $(\theta=\pi/2)$. (b) Energy spectrum of the emitted pairs for the $\gamma$-photon energy of $\omega'=50$ GeV: $\theta=0$ (thin blue line) and $\theta=\pi/2$ (green line). The vertical dash-dotted red (dashed black) lines show the location of third harmonic edge for $\theta = 0$ $(\theta=\pi/2)$. The laser parameters are $\omega = 4.65$ eV, $\xi = 0.4$. 
      }
        \label{NLBW}
  \end{center}
\end{figure}

In the applied scheme of Fig.~\ref{setP}, a particle would experience REF rather than a standing wave only if it 
propagates along the antinode plane (perpendicular to the beams axis), namely with $\theta = \pi/2$. On the other hand, we wish to detect the angle dependence of the effective mass. According to Fig.~\ref{effMassP}, this dependence is slow and monotonous. Thus, finding another configuration corresponding to $\theta=0$ may be sufficient. As explained above, the latter case is theoretically equivalent to a particle in the presence of a PWF. 
Hence, our reference configuration would be a $\gamma$- photon interacting with a circularly polarized PWF with the same $\xi$ value.
Since for the PWF the effective mass depends solely on $\xi$, the angle between the $\gamma$- photon and the laser may be chosen according to convenience.
In the following we assumed that this angle would be $\theta = \pi/2$. Namely, the reference configuration is identical to the one presented in Fig.~\ref{setP}, where only a single laser beam is active.

We start with the NLBW scattering. For this process to take place, the center of mass energy, $E_s = \sqrt{(sk + k')^2}=\sqrt{2s (k \cdot k')}$ should exceed $2 m_*$, where $s$ is the number of absorbed field photons and their wavevector reads $k = (\omega,0,0,0)$. This threshold suggests a simple way to measure the effective mass. Since for the set up illustrated in Fig. \ref{setP} we have $k \cdot k' = \omega \omega'$, the threshold energy for the incoming $\gamma$-photon is
\begin{equation}
\omega'_s= \frac{2m_*^2 }{s \omega }.
\label{eq:av gamma_th}
\end{equation}
Accordingly, increasing $\omega'$ for fixed laser parameters leads to a discrete change in the number of allowed channels in the vicinity of $\omega'_s$, leading to an abrupt jump in the total probability.
In order to detect this discontinuity two requirements should be fulfilled. First, the laser normalized amplitude should lay in the perturbative regime (i.e. $\xi \lesssim 1$), so that high harmonics are inhibited and the main contribution originates from the $s^{th}$ channel under consideration. Second, the threshold $\omega'_s$ should be remote from the sequential one $\omega'_{s+1}$, so that the influence of the 
$s^{th}$ channel would be distinguishable. Therefore, as Eq. (\ref{eq:av gamma_th}) implies, low harmonics are preferable. 
The total probability of pair production in dependence of the incoming $\gamma$-photon energy is shown in Fig.~\ref{NLBW}(a).
Since high $\omega'$ of $\gamma$-photon energies are difficult to achieve, we propose to increase $\omega$ by using  harmonics of the laser radiation, and consider the following laser parameters: $\xi = 0.4$, $\omega = 4.65$ eV, corresponding to the 3rd harmonic of Ti:S laser with intensity of $6 \times 10^{18}$ W/cm$^2$. As mentioned above, observing effective mass effects requires multi-cycle pulse. A 10 cycle pulse with the desired intensity focused on a spot with diameter of 10 wavelengths corresponds to $4$~mJ, which is realizable with the present laser technique \cite{wang}.
The $\gamma$-energies lay in the same GeV-range as those achieved in the E-144 experiment \cite{E144a, E144b}.
One may observe that the thresholds are $\omega'_2 = 65.2$~GeV for $\theta = 0$, and $60.5$~GeV for $\theta = \pi/2$, which using Eq.~(\ref{eq:av gamma_th}) correspond to $m_*(\theta=0)=m_*^P$ and $ m_*(\theta=\pi/2)=0.96m_*^P$, in accordance with Eq.~(\ref{eq:av effMass}). 
Notice that for $\theta=0 $ the quantum parameter is $\chi= \xi \omega \omega'/m^2$ whereas for $\theta = \pi/2 $ it reads $\xi \omega \omega'/m^2 \sin (\omega t)$. Accordingly, the average value of $\chi$ is lower in the second case and so is the corresponding rate.

Another indication to the effective mass may be observed in the spectrum of the created pair, as follows. 
From the energy momentum conservation $s k_{\mu} + \bar{P}_{\mu} = \bar{P}'_{\mu} + k'$, a restriction on the outcoming particles energy arises \cite{Supp}. For a given number of absorbed photons $s$, one may show that
\begin{equation}
\biggl|\varepsilon - \frac{\omega'}{2} \biggr| < \frac{\Delta_s}{2}, \quad \quad
\Delta_s = \omega' \sqrt{1 - \frac{s_0}{s}},
\label{eq:av BW_gap}
\end{equation}
where $s_0 =  2m_*^2/(\omega \omega')$.
As an example, the spectral probability associated with the created pair is depicted in Fig.~\ref{NLBW}b. The $\gamma$-photon energy is $50$ GeV and the laser parameters as described above. The widths of the 3rd harmonic are $\Delta_{3} = 0.34 \omega', 0.43 \omega'$ for $\theta = 0, \pi/2 $ respectively.
Employing (\ref{eq:av BW_gap}) one obtained the same effective mass values written above.

Furthermore, the effective mass is manifested in the NLC spectra (the PWF case was discussed in \cite{Harvey2}).
A straightforward kinematic calculation \cite{Supp}, shows that for a given $s$, the emitted photon has a cutoff energy, known as "edge"
\begin{equation}
\omega'_e = \frac{s \omega \varepsilon}{\varepsilon (1-\bar{\upsilon})+ s \omega}, 
\quad \quad
\bar{\upsilon} = \frac{p}{\sqrt{m_*^2+p^2}},
\label{eq:av edge1}
\end{equation}
where $\bar{\upsilon}$ is the absolute value of the cycle-averaged velocity.
As a result, the effective mass affects the edge location. 
In principle, since the effective mass is momentum-dependent (as shown in Fig.~\ref{effMassP}), it may differ for the incoming and outcoming particles. We study the process in the classical regime, $\chi \ll 1$, because 
the regime where both $\xi \sim 1$ and $\chi \sim 1$ are fulfilled would require very high frequency colliding laser beams (with photon energies of MeV range, which is beyond contemporary experimental reach).
As a result, the recoil is negligible ($p \approx p'$) and therefore effective mass of the incoming and outcoming electron are the same.
The emission properties of a particle propagating in the electric field plane may be measured in a set up similar to the one in Fig. \ref{setP}, where the $\gamma$-photons are replaced by high energy electrons. 
A notable fact is that the NLC process has no threshold, as opposed to the NLBW discussed above. Namely, all possible channels $s$ are allowed, regardless of the incoming electron energy. Consequently, the first harmonic of the Ti:S laser as well as a modest electron energy are sufficient. Furthermore, the NLC spectrum is less sensitive to increase in the field amplitude $\xi$ as compared to the NLBW one. As a result, one may use higher values of $\xi$ without losing the edge structure.
These two facts allow one to explore the edge structure and thus the effective mass for $p \sim m \xi$. 
In this regime, as opposed to the $p \gg m \xi$ case discussed above, the effective mass depends not only on $\theta$ but on $p$ as well (see Fig.~\ref{effMassP}).

\begin{figure}
  \begin{center}     	   \includegraphics[width=0.45\textwidth]{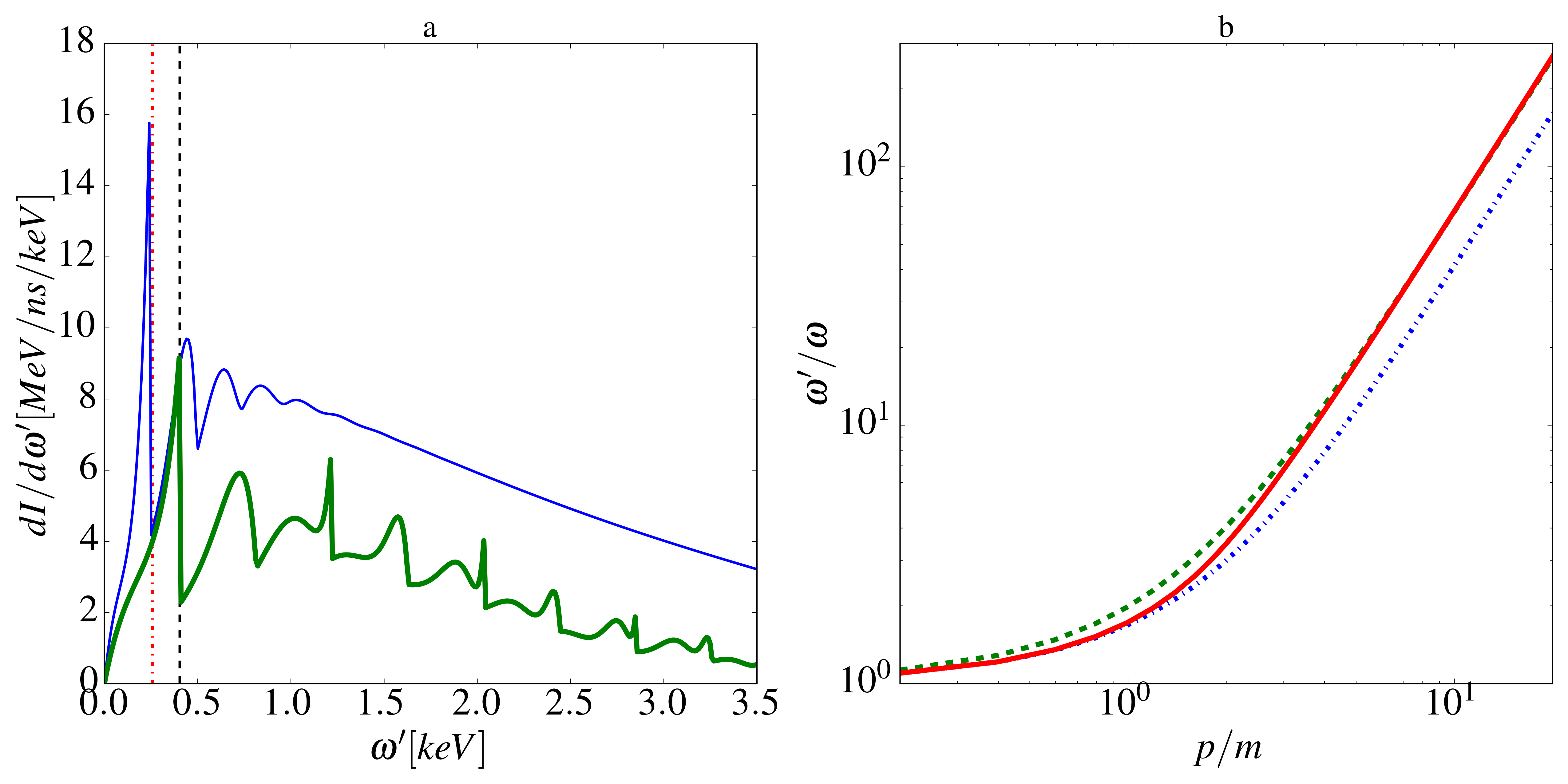}
      \caption{(a)  NLC emission spectrum:  $\theta=0$ (thin blue line) and $\theta=\pi/2$ (green line). Simulation parameters: $\omega = 1.55$ eV, $\xi = 2$, $p/m=20$. The vertical dash-dotted red (dashed black) line shows the edge location of the first harmonic $\theta = 0$ $(\theta=\pi/2)$. (b) The edge energy for $\theta=\pi/2$, normalized by $ \omega$, as a function of $p/m$ (red line). As a reference, the prediction of Eq. (\ref{eq:av edge1}) for the limiting cases $m_*(p=0)$ (dash-dotted blue line) and $m_*(p \gg m \xi)$ (dashed green line).   }
        \label{NLC}
  \end{center}
\end{figure}
Figure~\ref{NLC}(a) presents the NLC spectrum for $\theta = 0$ and $\theta=\pi/2$. The laser frequency and intensity are $\omega =1.6$ eV, $I = 1.7 \times 10^{19}$W/cm$^2$, corresponding to $\xi = 2$. The particle initial momentum is $p/m=20$. One may see that the harmonics edges become smeared with increasing $s$. As a result, it is convenient to take a closer look at the first harmonic only. The edge locations corresponding to $\theta=0$ and $\theta=\pi/2$ are $\omega'_e =0.26 $ keV and  $\omega'_e = 0.4$ keV, respectively. From Eq.~(\ref{eq:av edge1}) one may calculate the effective masses $m_*(\theta=0)=m_*^P, m_*(\theta=\pi/2)=0.77m_*^P$, in agreement with the prediction of Eq. (\ref{eq:av effMass}).
As in the NLBW case, the average value of $\chi$ is smaller for $\theta = \pi/2$, leading to a lower spectrum.
By varying the incoming electron momentum one may observe the shift of the edge location. The shift of the edge from the calculated spectra is summarized in Fig.~\ref{NLC}(b). From the latter the effective mass is deduced using Eq.~(\ref{eq:av edge1}), which is in accordance with the function $m_*(\pi/2, p/m)$ presented in Fig.~\ref{effMassP}(a) and calculated from the classical trajectory.  
As a reference, the prediction of Eq. (\ref{eq:av edge1}) for constant effective mass values corresponding to the limiting cases $p=0$ and $p \gg m \xi$ are shown in Fig.~\ref{eq:av edge1}(b) as well. As expected, the curve obtained from the edge location shift (solid red line) interpolates continuously between the two other ones. 

Concluding, the emergence of a momentum-dependent effective mass in the presence of a strong REF has been demonstrated. As a result, the pair production threshold by a $\gamma$-photon and the harmonic edges in the pair spectrum, depend on its angle with respect of the field plane. Moreover, from the edges of the photon emission spectrum of an electron in the presence of this field, the momentum-dependence of the effective mass could be measured. These predictions may be put to the test with present day facilities.

ER is grateful to Q. Z. Lyu for helpful discussions.
ER acknowledges support from the Alexander von Humboldt Foundation.

\clearpage


\section*{Supplementary material}

# Supplemental Materials
## to the paper "Nonlinear QED in an ultrastrong rotating electric field: Signatures of the momentum-dependent effective mass"

## I. TRAJECTORY

In the semiclassical formalism, employed in this paper, the classical trajectory of the particle is the cornerstone of the rate calculation. Let us calculate explicitly the trajectory of an electron moving in a rotating electric field (REF), described by a vector potential $\mathbf{A}(t) = a(\cos\omega t, \sin\omega t, 0)$. The particle location is given by

$$\mathbf{x}(t) = \int_{t_0}^{t} dt' \boldsymbol{v}(t') = \int_{t_0}^{t} dt' \frac{\mathbf{P}(t')}{\mathcal{E}(t')}. \tag{1}$$

Due to the canonical momentum conservation, the kinetic momentum is straightforwardly derived

$$\mathbf{P}(t) = \mathbf{p} + e\mathbf{A}(t), \tag{2}$$

and using the dispersion relationship $\mathcal{E}(t) = \sqrt{\mathbf{P}^2(t) + m^2}$, one arrives at

$$\mathcal{E}(t) = \sqrt{\varepsilon_0^2 + (ea)^2 + 2eap|\sin\theta|\cos(\omega t - \nu)}, \tag{3}$$

where $\tan\nu = p_y/p_x$. It may be represented as

$$\mathcal{E} = G\sqrt{1 - \mu\sin^2\left(\frac{\omega t - \nu}{2}\right)}, \tag{4}$$

where the following quantities are introduced

$$G \equiv \sqrt{m^2(1+\xi^2)p^2 + 2m\xi p|\sin\theta|}, \tag{5}$$

and

$$\mu \equiv \frac{4m\xi p|\sin\theta|}{G^2}. \tag{6}$$

Substituting the explicit expression for the energy into Eq. (1), one obtains the particle coordinate. Its $x$ component reads

$$x(t) = \frac{1}{G}\int_{t_0}^{t} dt' \left[\frac{p_x}{\sqrt{1 - \mu\sin^2\left(\frac{\omega t'-\nu}{2}\right)}} + \frac{ea\cos(\omega t')}{\sqrt{1 - \mu\sin^2\left(\frac{\omega t'-\nu}{2}\right)}}\right]. \tag{7}$$

A variable change $\phi \equiv (\omega t - \nu)/2$ yields

$$x(t) = \frac{2}{\omega G}\int_{\phi_0}^{\phi} d\phi' \left[\frac{p_x}{\sqrt{1-\mu\sin^2\phi'}} + \frac{ea\cos(2\phi'+\nu)}{\sqrt{1-\mu\sin^2\phi'}}\right]. \tag{8}$$

The latter takes the form

$$x(t) = \frac{2}{\omega G}\left[p_x\mathcal{J}_1 + m\xi(\cos\nu\mathcal{J}_2 - \sin\nu\mathcal{J}_3)\right], \tag{9}$$

where the following integrals are defined:

$$\mathcal{J}_1(x|\mu) \equiv \int_0^x dx' \frac{1}{\sqrt{1-\mu\sin^2 x'}}, \tag{10}$$

$$\mathcal{J}_2(x|\mu) \equiv \int_0^x dx' \frac{\cos(2x')}{\sqrt{1-\mu\sin^2 x'}}, \tag{11}$$

$$\mathcal{J}_3(x|\mu) \equiv \int_0^x dx' \frac{\sin(2x')}{\sqrt{1-\mu\sin^2 x'}}, \tag{12}$$

and $x = \omega t/2$. These integrals admit analytical solution

$$\mathcal{J}_1(x|\mu) = E_1(x|\mu), \tag{13}$$

$$\mathcal{J}_2(x|\mu) = \frac{(\mu-2)E_1(x|\mu) + 2E_2(x|\mu)}{\mu}, \tag{14}$$

$$\mathcal{J}_3(x|\mu) = \frac{2}{\mu}\left[1 - \sqrt{1-\mu\sin^2 x}\right], \tag{15}$$

where $E_1(x|\mu), E_2(x|\mu)$ are the incomplete elliptic integral of first and second kind, respectively

$$E_1(x|\mu) \equiv \int_0^x dx' \frac{1}{\sqrt{1-\mu\sin^2 x'}}, \tag{16}$$

$$E_2(x|\mu) \equiv \int_0^x dx' \sqrt{1-\mu\sin^2 x'}. \tag{17}$$

Analogously, for the $y$ component of the coordinate one obtains

$$y(t) = \frac{2}{\omega G}\int_{\phi_0}^{\phi} d\phi' \left[\frac{p_y}{\sqrt{1-\mu\sin^2\phi'}} + \frac{ea\sin(2\phi'+\nu)}{\sqrt{1-\mu\sin^2\phi'}}\right], \tag{18}$$

which reads

$$y(t) = \frac{2}{\omega G}\left[p_y\mathcal{J}_1 + m\xi(\sin\nu\mathcal{J}_2 + \cos\nu\mathcal{J}_3)\right]. \tag{19}$$

The vector potential has no $z$ component, so for $z(t)$ we simply have

$$z(t) = \frac{2}{\omega G} p_z \mathcal{J}_1. \tag{20}$$

In addition to the trajectory, the average velocity is required as well for the purpose of rate calculation. Applying the definition $\bar{v}_x \equiv [x(T) - x(0)]/T$, where $T = 2\pi/\omega$, we obtain

$$\bar{v}_x = \frac{1}{\pi G}\left[p_x\mathcal{J}_1 + m\xi(\cos\nu\mathcal{J}_2 - \sin\nu\mathcal{J}_3)\right]\Big|_0^\pi. \tag{21}$$

One may easily find

$$\mathcal{J}_1(x|\mu)\Big|_0^\pi = 2E_1(\mu), \tag{22}$$

$$\mathcal{J}_2(x|\mu)\Big|_0^\pi = \frac{2(\mu-2)E_1(\mu) + 4E_2(\mu)}{\mu}, \tag{23}$$

$$\mathcal{J}_3(x|\mu)\Big|_0^\pi = 0, \tag{24}$$

where the complete elliptic integrals are given by $E_1(\mu) = E_1(\frac{\pi}{2}|\mu)$, $E_2(\mu) = E_2(\frac{\pi}{2}|\mu)$. Accordingly, the $x$ component of the average velocity takes the form

$$\bar{v}_x = \frac{2}{\pi G}\left[p_x E_1(\mu) + m\xi \cos\nu\left(\frac{2E_2(\mu) + (\mu-2)E_1(\mu)}{\mu}\right)\right]. \tag{25}$$

Similarly, the other components are given by

$$\bar{v}_y = \frac{2}{\pi G}\left[p_y E_1(\mu) + m\xi \sin\nu\left(\frac{2E_2(\mu) + (\mu-2)E_1(\mu)}{\mu}\right)\right] \tag{26}$$

$$\bar{v}_z = \frac{2}{\pi G} p_z E_1(\mu). \tag{27}$$

## II. APPROXIMATED EFFECTIVE MASS

As explained in the main text, the effective mass is defined as $\sqrt{\bar{P}^2}$. Since $\bar{\mathbf{P}} = \mathbf{p}$ we have

$$m_* = \sqrt{\bar{\mathcal{E}}^2 - \mathbf{p}^2}. \tag{28}$$

The cycle-averaged energy is given by

$$\bar{\mathcal{E}} = \frac{2}{\pi} G E_2(\mu), \tag{29}$$

In the following we would like to Taylor expand the effective mass for $p \gg m\xi$. As we show below, the first order vanishes and, therefore, we evaluate it up to second order. We introduce the following quantities

$$\delta \equiv \frac{4m\xi p |\sin\theta|}{R^2}, \quad R \equiv \sqrt{m^2\xi^2 + m^2 + p^2}. \tag{30}$$

$G, \mu$ defined above read in terms of these variables:

$$G = R\sqrt{1 + \frac{\delta}{2}}, \tag{31}$$

$$\mu = \frac{\delta}{1 + \delta/2}. \tag{32}$$

Substituting Eqs.(32) and (31) into Eq. (29), one finds

$$\bar{\mathcal{E}} = \frac{2R}{\pi}\sqrt{1 + \frac{\delta}{2}} E_2\left(\frac{\delta}{1 + \delta/2}\right). \tag{33}$$

One may notice that up to third order

$$\delta \approx 4|\sin\theta|\left(\frac{m\xi}{p}\right) + O\left(\left[\frac{m\xi}{p}\right]^3\right). \tag{34}$$

As a result, up to second order, we may expand with respect to $\delta$ instead of $m\xi/p$. Employing the following Taylor expansions

$$E_2(x) \approx \frac{\pi}{2}\left(1 - \frac{x}{4} - \frac{3x^2}{64}\right), \tag{35}$$

$$\sqrt{1+x} \approx 1 + \frac{x}{2} - \frac{x^2}{8}, \tag{36}$$

and substituting Eq. (33) into Eq. (28), one obtains

$$m_*^2 = (m_*^P)^2 - \frac{\delta^2 R^2}{32}, \tag{37}$$

where $m_*^P = m\sqrt{1+\xi^2}$. Since

$$\delta^2 R^2 = \frac{16 m^2 p^2 \xi^2 \sin^2\theta}{R^2} \approx 16 m^2 \xi^2 \sin^2\theta, \tag{38}$$

one may see that Eq. (37) becomes

$$\frac{m_*}{m_*^P} = \sqrt{1 - \frac{\xi^2}{2(1+\xi^2)}\sin^2\theta}. \tag{39}$$

## III. NONLINEAR COMPTON SCATTERING

### A. Probability

As demonstrated in [1], under the condition $\varepsilon \gg m\xi$, the quantum and semiclassical [2, 3] approaches coincide. Since the latter allows for simpler calculation, it will be used in this work. According to this approximation, the probability of a Dirac particle to emit a photon with a four-momentum $k'$ is given by

$$d\mathcal{P} = \frac{\alpha}{(2\pi)^2} d^3\mathbf{k}'$$
$$\int_{-\infty}^\infty dt_1 \int_{-\infty}^\infty dt_2 N_{21} \exp\left[i\frac{\varepsilon}{\varepsilon - \omega'} k' \cdot [x(t_1) - x(t_2)]\right], \tag{40}$$

where $x_\mu = [t, \mathbf{x}(t)]$ and $\mathbf{x}(t)$ is the particle classical trajectory found above. The prefactor is given by

$$N_{21} \equiv \frac{1}{2\varepsilon'^2}\left[\left(\varepsilon'^2 + \varepsilon^2\right)[\boldsymbol{v}(t_1)\cdot\boldsymbol{v}(t_2) - 1] + \frac{\omega'^2 m^2}{\varepsilon^2}\right]. \tag{41}$$

It should be mentioned that this expression already contains averaging over the incoming electron spin and summing over the outgoing electron (photon) spin (polarization), respectively. Since $\boldsymbol{v}(t_1)\boldsymbol{v}(t_2) - 1 = -v_\mu(t_1)v^\mu(t_2)$, Eq. (40) can be cast in the following form

$$d\mathcal{P} = \frac{\alpha}{(2\pi)^2}|\mathcal{K}|^2 d^3 k', \tag{42}$$

where

$$|\mathcal{K}|^2 \equiv -\left(\frac{\varepsilon'^2 + \varepsilon^2}{2\varepsilon'^2}\right)|\mathcal{T}_\mu|^2 + \frac{m^2\omega'^2}{2\varepsilon'^2\varepsilon^2}|\mathcal{I}|^2, \tag{43}$$

with

$$\mathcal{I} \equiv \int_{-\infty}^{\infty} dt e^{i\psi}, \quad \mathcal{T}_\mu \equiv \int_{-\infty}^{\infty} dt v_\mu(t) e^{i\psi}, \quad (44)$$

and

$$\psi = \frac{\varepsilon}{\varepsilon'} k' \cdot x(t) = \frac{\varepsilon \omega' t}{\varepsilon'} (1 - \mathbf{x} \cdot \mathbf{n}'), \quad (45)$$

where $k'_\mu = \omega'(1, \mathbf{n}')$. Since we are dealing with a periodic motion, the phase may be decomposed to a periodic and non-periodic parts, $\psi = \psi_p + \psi_{np}\omega t$ with

$$\psi_p = \frac{\varepsilon}{\varepsilon'} \mathbf{n}' \cdot \mathbf{x}_p, \quad \psi_{np} = \frac{\varepsilon \omega'}{\varepsilon'} (1 - \bar{\mathbf{v}} \cdot \mathbf{n}'), \quad (46)$$

where the periodic part of the trajectory is given by

$$\mathbf{x}_p(t) = \mathbf{x}(t) - \bar{\mathbf{v}} t. \quad (47)$$

Since we assume that the incoming electron propagates along the $x$-axis, $p_\mu = (\varepsilon, p_x, 0, 0)$, the emitted photon parametrization is defined accordingly

$$\mathbf{n}' = (\cos\theta_e, \sin\theta_e \sin\varphi_e, \sin\theta_e \cos\varphi_e), \quad (48)$$

where $\theta_e, \varphi_e$ are the polar and azimuthal angles with respect to the $x$-axis, respectively. Replacing the periodic part of the integrands by their Fourier series, the integrals are solved

$$\mathcal{T}^\mu = 2\pi \sum_s \mathcal{T}_s^\mu \delta(\Omega_s), \quad \mathcal{I} = 2\pi \sum_s \mathcal{I}_s \delta(\Omega_s), \quad (49)$$

where the argument of the delta function reads

$$\Omega_s \equiv \psi_{np} - s\omega, \quad (50)$$

and the Fourier coefficients are

$$\mathcal{T}_s^\mu = \frac{1}{T} \int_0^T dt v_\mu(t) e^{i(s\omega t - \psi_p)}, \quad (51)$$

$$\mathcal{I}_s = \frac{1}{T} \int_0^T dt e^{i(s\omega t - \psi_p)}. \quad (52)$$

With the aid of the condition $\Omega_s = 0$, forced by the delta functions, the angle $\theta_e$ is found

$$\cos\theta_e = \frac{1}{\bar{v}}\left(1 - \frac{s\omega\varepsilon'}{\omega'\varepsilon}\right). \quad (53)$$

Using Eq. (48), the periodic part of the phase takes the form

$$\psi_p = \frac{\varepsilon\omega'}{\varepsilon - \omega'}\left[\cos\theta_e x_p(t) + \sin\theta_e \sin\varphi_e y_p(t)\right]. \quad (54)$$

Substituting Eq. (49) into Eq. (43), and using the identity $\delta^2(\Omega_s) = \frac{\tau}{2\pi}\delta(\Omega_s)$, with the interaction time $\tau$, one obtains

$$|\mathcal{K}|^2 = 2\pi \sum_s \mathcal{K}_s^2 \delta(\Omega_s)\tau, \quad (55)$$

where

$$|\mathcal{K}_s|^2 = -\left(\frac{\varepsilon'^2 + \varepsilon^2}{2\varepsilon'^2}\right)|\mathcal{T}_s^\mu|^2 + \frac{m^2 \omega'^2}{2\varepsilon'^2 \varepsilon^2}|\mathcal{I}_s|^2. \quad (56)$$

Using $d^3\mathbf{k}' = \omega'^2 d(\cos\theta_e) d\varphi_e$ and integrating Eq. (42) over $\cos\theta_e$ yields

$$\frac{dI}{d\omega' d\varphi_e} = \frac{1}{(2\pi)}\omega'^2 \sum_s |\mathcal{K}_s|^2 \left|\frac{d\Omega_s}{d(\cos\theta_e)}\right|^{-1}, \quad (57)$$

where the relation $dI = \omega' d\mathcal{P}/\tau$ between the probability and the radiation intensity was employed. Since $\bar{\mathbf{v}} \cdot \mathbf{n}' = \bar{v}\cos\theta_e$, from Eq.(50) it follows that

$$\left|\frac{d\Omega_s}{d(\cos\theta_e)}\right| = \frac{\varepsilon\omega'\bar{v}}{\varepsilon'}. \quad (58)$$

Hence, the final expression takes the form

$$\frac{dI}{d\omega' d\varphi_e} = \frac{\alpha\omega'}{4\pi\varepsilon^3\varepsilon'} \sum_s \left[-\varepsilon^2\left(\varepsilon^2 + \varepsilon'^2\right)|\mathcal{T}_s^\mu|^2 + m^2\omega'^2|\mathcal{I}_s|^2\right]. \quad (59)$$

### B. Kinematics

The highest possible value of $\omega'$ associated with a given harmonics $s$ may be derived from kinematic considerations. Using $\varepsilon' = \varepsilon - \omega'$, the emitted photon energy stems from the kinematic relation Eq. (53)

$$\omega' = \frac{s\omega\varepsilon}{\varepsilon(1 - \bar{v}\cos\theta_e) + s\omega}. \quad (60)$$

The maximal value of $\omega'$ corresponds to $\cos\theta_e = 1$. This result may be derived by an alternative kinematic approach. The energy momentum conservation of this process reads

$$\bar{P}_\mu + sk_\mu = \bar{P}'_\mu + k'_\mu, \quad (61)$$

where $k_\mu = (\omega, 0, 0, 0)$. Therefore, the spatial momentum conservation yields

$$p'_\parallel + k'_\parallel = p, \quad p'_\perp = -k'_\perp, \quad (62)$$

where $\parallel$ and $\perp$ designate the parallel and transverse components of the momenta with respect to the incoming particle direction, respectively. Then the total outgoing momentum reads

$$p' = \sqrt{m^2 + p_\parallel'^2 + p_\perp'^2} = \sqrt{\omega'^2 + p^2 - 2p\omega'\cos\theta_e}, \quad (63)$$

where $k'_\parallel = \omega'\cos\theta_e$ and $k'_\perp = \omega'\sin\theta_e$. Substituting $p'$ into the energy conservation equation

$$\bar{\mathcal{E}} + s\omega = \omega' + \sqrt{m_*^2 + p'^2}, \quad (64)$$

one obtains the energy of the emitted photon

$$\omega' = \frac{2s\omega\bar{\mathcal{E}} + s^2\omega^2}{2(\bar{\mathcal{E}} - p\cos\theta_e + s\omega)}. \quad (65)$$

Recalling that $\bar{\mathbf{P}} = \mathbf{p}$, the absolute value of the average velocity is given by $\bar{v} = p/\bar{\mathcal{E}}$. As a result we have

$$\omega' = \frac{s\omega\bar{\mathcal{E}}}{\bar{\mathcal{E}}(1 - \bar{v}\cos\theta_e) + s\omega}. \quad (66)$$

where $s\omega \ll \bar{\mathcal{E}}$ was assumed. Approximating $\bar{\mathcal{E}} \approx \varepsilon$ one returns to Eq. (60) given above.

## IV. NONLINEAR BREIT-WHEELER PROCESS

### A. Probability

Owing to the crossing symmetry relating the NLC and NLBW processes [3, 4], the semiclassical probability associated with the latter takes the form

$$d\mathcal{P} = \frac{\alpha}{(2\pi)^2 \omega'} |\mathcal{K}|^2 d^3\mathbf{p}. \tag{67}$$

The difference with respect to the photon emission expression of Eq. (42) is the outgoing particle phase space, namely $d^3\mathbf{k}' \to d^3\mathbf{p}$. Therefore, the final result Eq. (59) should be only multiplied by a factor $\varepsilon^2/\omega'^2$. Moreover, in this case we are interested in the emission rate rather than intensity, leading to additional $1/\omega'$ factor. Finally, one obtains

$$\frac{dW}{d\varepsilon d\varphi_e} = \frac{\alpha}{4\pi\varepsilon\varepsilon'\omega'^2} \sum_s [-\varepsilon^2 (\varepsilon^2 + \varepsilon'^2) |\mathcal{T}_s^\mu|^2 + m^2 \omega'^2 |\mathcal{I}_s|^2], \tag{68}$$

where $\mathcal{T}_s^\mu, \mathcal{I}_s$ are given by Eqs. (51) and (52). Another modification with respect to NLC scattering lies in the periodic part of the phase. Since the incoming photon propagates along the $x$-axis, one may write $\mathbf{n}' = (1, 0, 0)$. Therefore, the outgoing electron four-momentum is parameterized as

$$p_\mu = (\varepsilon, p\cos\theta_e, p\sin\theta_e \sin\varphi_e, p\sin\theta_e \cos\varphi_e). \tag{69}$$

Accordingly, $\psi_p$ may be written as

$$\psi_p = \frac{\varepsilon\omega'}{\omega' - \varepsilon} \cos\theta_e x_p(t), \tag{70}$$

where the trajectory is given by Eq. (9), and the relation $\varepsilon' = \omega' - \varepsilon$ is used.

### B. Kinematics

As in the NLC case, the effective mass may be inferred from the maximal value of the outgoing particle energy for a given harmonic $s$. The kinematic relation Eq. (53), together with $\varepsilon' = \omega' - \varepsilon$, yields

$$\cos\theta_e = \frac{1}{\bar{v}} \left[1 - \frac{s\omega(\omega' - \varepsilon)}{\omega' \varepsilon}\right]. \tag{71}$$

Since $\cos\theta_e \leq 1$ and employing

$$1 - \bar{v} = \frac{\varepsilon - \sqrt{\varepsilon^2 - m_*^2}}{\varepsilon} \approx \frac{m_*^2}{2\varepsilon^2}, \tag{72}$$

one obtains

$$\frac{s\omega(\omega' - \varepsilon)}{\omega' \varepsilon} \geq \frac{m_*^2}{2\varepsilon^2}. \tag{73}$$

Solving the quadratic inequality for $\varepsilon$ one arrives at

$$\left|\varepsilon - \frac{\omega'}{2}\right| \leq \sqrt{1 - \frac{s_0}{s}}, \tag{74}$$

where $s_0 = 2m_*^2/(\omega\omega')$.



\end{document}